\documentclass[prl,twocolumn,superscriptaddress]{revtex4-1}
\usepackage{}
\usepackage{amssymb}
\usepackage{amsfonts}
\usepackage{bbm}
\usepackage{mathrsfs}
\usepackage{graphics,graphicx,epsfig,bm,amsmath,amsthm,amssymb}
\usepackage{bm}
\usepackage{bbm}
\usepackage{longtable}
\usepackage{multirow}
\usepackage{array}
\usepackage{color}
\usepackage[usenames,dvipsnames]{xcolor}
\newcommand{\ket}[1]{\left| #1\right\rangle}      

\usepackage{float}
\usepackage[a4paper,colorlinks=true,
linkcolor=blue,citecolor=blue,
pdfauthor={ },
pdftitle={ },
pdfsubject={ },
pdfkeywords={ }]{hyperref}

\begin{document}

\title{Searching for new particles beyond the standard model with a single electron-spin quantum sensor}

\author{Xing Rong}
\thanks{These authors contributed equally to this work.}
\affiliation{CAS Key Laboratory of Microscale Magnetic Resonance and Department of Modern Physics, University of Science and Technology of China (USTC), Hefei 230026, China}
\affiliation{Hefei National Laboratory for Physical Sciences at the Microscale, USTC}
\affiliation{Synergetic Innovation Center of Quantum Information and Quantum Physics, USTC}
\author{Mengqi Wang}
\thanks{These authors contributed equally to this work.}
\affiliation{CAS Key Laboratory of Microscale Magnetic Resonance and Department of Modern Physics, University of Science and Technology of China (USTC), Hefei 230026, China}
\affiliation{Hefei National Laboratory for Physical Sciences at the Microscale, USTC}
\author{Jianpei Geng}
\thanks{These authors contributed equally to this work.}
\affiliation{CAS Key Laboratory of Microscale Magnetic Resonance and Department of Modern Physics, University of Science and Technology of China (USTC), Hefei 230026, China}
\affiliation{Hefei National Laboratory for Physical Sciences at the Microscale, USTC}
\author{Xi Qin}
\affiliation{CAS Key Laboratory of Microscale Magnetic Resonance and Department of Modern Physics, University of Science and Technology of China (USTC), Hefei 230026, China}
\affiliation{Hefei National Laboratory for Physical Sciences at the Microscale, USTC}
\affiliation{Synergetic Innovation Center of Quantum Information and Quantum Physics, USTC}
\author{Maosen Guo}
\affiliation{CAS Key Laboratory of Microscale Magnetic Resonance and Department of Modern Physics, University of Science and Technology of China (USTC), Hefei 230026, China}
\author{Man Jiao}
\affiliation{CAS Key Laboratory of Microscale Magnetic Resonance and Department of Modern Physics, University of Science and Technology of China (USTC), Hefei 230026, China}
\author{Yijin Xie}
\affiliation{CAS Key Laboratory of Microscale Magnetic Resonance and Department of Modern Physics, University of Science and Technology of China (USTC), Hefei 230026, China}
\author{Pengfei Wang}
\affiliation{CAS Key Laboratory of Microscale Magnetic Resonance and Department of Modern Physics, University of Science and Technology of China (USTC), Hefei 230026, China}
\affiliation{Hefei National Laboratory for Physical Sciences at the Microscale, USTC}
\affiliation{Synergetic Innovation Center of Quantum Information and Quantum Physics, USTC}
\author{Pu Huang}
\affiliation{CAS Key Laboratory of Microscale Magnetic Resonance and Department of Modern Physics, University of Science and Technology of China (USTC), Hefei 230026, China}
\affiliation{Hefei National Laboratory for Physical Sciences at the Microscale, USTC}
\affiliation{Synergetic Innovation Center of Quantum Information and Quantum Physics, USTC}
\author{Fazhan Shi}
\affiliation{CAS Key Laboratory of Microscale Magnetic Resonance and Department of Modern Physics, University of Science and Technology of China (USTC), Hefei 230026, China}
\affiliation{Hefei National Laboratory for Physical Sciences at the Microscale, USTC}
\affiliation{Synergetic Innovation Center of Quantum Information and Quantum Physics, USTC}
\author{Yi-Fu Cai}
\affiliation{CAS Key Laboratory for Research in Galaxies and Cosmology, Department of Astronomy, USTC}
\affiliation{School of Astronomy and Space Science, USTC}
\author{Chongwen Zou}
\affiliation{National Synchrotron Radiation Laboratory, USTC}
\author{Jiangfeng Du}
\email{djf@ustc.edu.cn}
\affiliation{CAS Key Laboratory of Microscale Magnetic Resonance and Department of Modern Physics, University of Science and Technology of China (USTC), Hefei 230026, China}
\affiliation{Hefei National Laboratory for Physical Sciences at the Microscale, USTC}
\affiliation{Synergetic Innovation Center of Quantum Information and Quantum Physics, USTC}

\begin{abstract}

%
Searching for new particles beyond the standard model is crucial for understanding several fundamental conundrums in physics and astrophysics.
Amongst them, axions or similar hypothetical pseudoscalar bosons would mediate electron-nucleon interactions.
While previous experiments set stringent upper bounds of this interaction strength with force range over $20 ~\mu$m, experimental searching at shorter force range remains elusive.
We develop a method that utilizes a near-surface Nitrogen-vacancy center as a quantum sensor to explore such interaction.
New constraints for axion-mediated electron-nucleon coupling, $g_s^Ng_p^e$, have been set for the force range $0.1 $--$23~\mu$m.
The obtained upper bound of the interaction at $20~\mu$m, $g_s^Ng_p^e < 6.24\times10^{-15}$,  is two orders of magnitude more stringent than that set by earlier experiments.
Our method can be further extended to investigate other spin-dependent interactions and opens the door for the single-spin quantum sensor to explore new physics beyond the standard model.
%

\end{abstract}

\maketitle
Given our ignorance of the ultraviolet completion of particle physics, it is important to develop techniques to search for new particles beyond the standard model \cite{RMP_ExtSM}.
Among them, one type of ultralight scalars, such as axions and axionlike particles (ALPs) \cite{Arxiv_ALP}, has attracted a lot of attention in a wide variety of researches.
This has been well motivated for decades from the need of cosmology\cite{PR_axion_cosmology}, namely, the dark matter candidate\cite{PhysRep_DM} , the dark energy candidate\cite{PRL_DE}, and from the understanding of the symmetries of charge conjugation and parity in quantum chromodynamics (QCD)\cite{PRL_PQ} as well as predictions from fundamental theories such as string theory \cite{RMP_ExtSM}.
%
The exchange of such particles results in spin-dependent forces, which were originally investigated by Moody and Wilczek \cite{PRD_NewMacroForce}.
Various laboratory ALP searching experiments focus on the detection of macroscopic monopole-dipole forces between polarized electrons/nucleons and unpolarized nucleons \cite{PRL67_1735, PRL77_2170, PRL82_2439, PRD78_092006, PRL105_170401, PRL106_041801, PRL111_102001}.
Previous laboratory searching has set the limit on the monopole-dipole coupling between electron and nucleon, $g_s^Ng_p^e$,  with a force range $\lambda >  20~\mu$m\cite{PRD_LimitCoupling}.
The experimental investigation of this interaction at force range shorter than $ 20~\mu$m, however, remains elusive due to the following challenges.
The size of the sensor should be small compared to the micrometer force range.
The geometry of the sensor should allow close proximity between the electron and the nucleon.
The sensitivity of the sensor should be sufficient good for searching for or  providing stringent bound for such interaction.
The unwanted noises, such as the magnetic and electric field introduced by environment, should be isolated well.


We use a near-surface single electron spin, which is a Nitrogen vacancy (NV) center in diamond, to investigate the monopole-dipole interaction between a electron spin and nucleons.
The axion-mediated monopole-dipole interaction can be described as \cite{JHighEnergyPhys_MacroForce}
\begin{equation}
\label{Vsp}
V_{\textrm{sp}}(\vec{r})=\frac{\hbar^2g_s^Ng_p^e}{8\pi m}(\frac{1}{\lambda r}+\frac{1}{r^2})e^{-\frac{r}{\lambda}}\vec{\sigma}\cdot\hat{r},
\end{equation}
where $\vec{r}$ is the displacement vector between the electron and nucleon, $r=|\vec{r}|$ and $\hat{r}=\vec{r}/r$ are the displacement and the unit displacement vector, $g_s^N$ and $g_p^e$ are the scalar and pseudoscalar coupling constants of the ALP to the nucleon and to the electron, $m$ is mass of the electron, $\lambda=\hbar/(m_ac)$  is the force range, $m_a$ is the mass of the ALP, $\vec{\sigma}$ is the Pauli vector of the electron spin, $\hbar$ is Plank's constant divided by $2\pi$, and $c$ is the speed of light.
Such interaction is equivalent to the Hamiltonian of the electron spin in an effective magnetic field $\vec{B}_{\textrm{sp}}(\vec{r})$ arising from the nucleon,
\begin{equation}
\label{Bsp}
\vec{B}_{\textrm{sp}}(\vec{r})=\frac{\hbar g_s^Ng_p^e}{4\pi m\gamma}(\frac{1}{\lambda r}+\frac{1}{r^2})e^{-\frac{r}{\lambda}}\hat{r},
\end{equation}
where $\gamma$ is the gyromagnetic ratio of the electron spin.

\begin{figure*}[http]
\centering
\includegraphics[width=1.5\columnwidth]{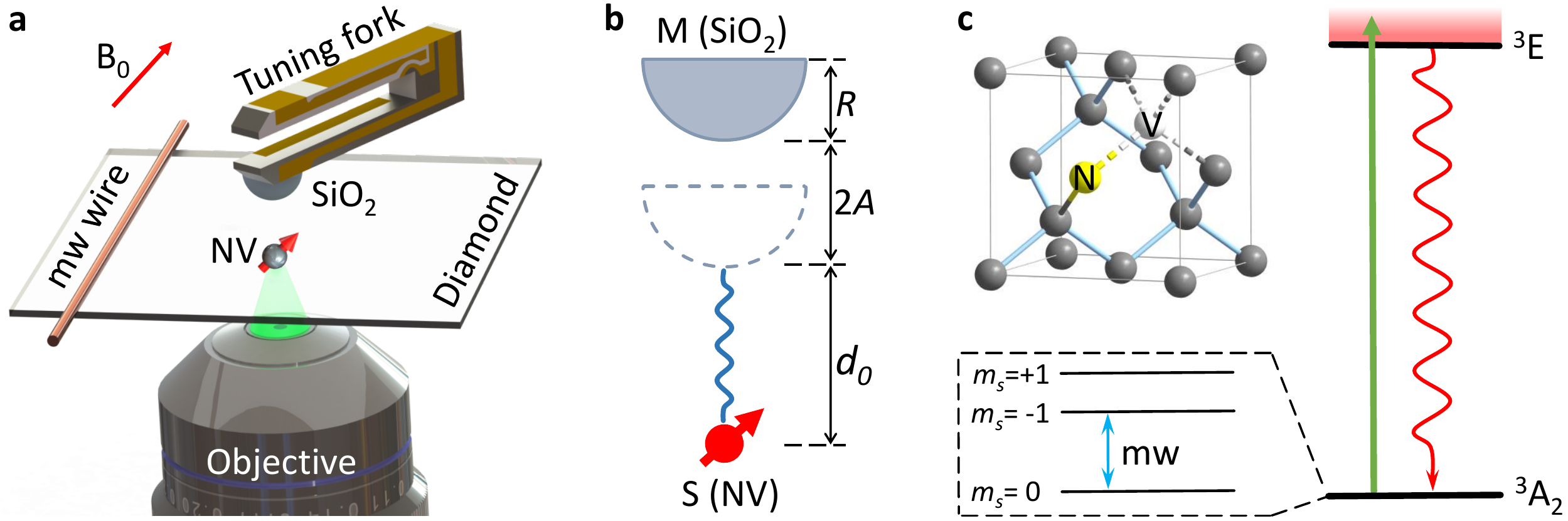}
    \caption{Experimental setup and the quantum sensor. (\textbf{a}) Schematic experimental setup. A nitrogen-vacancy center in diamond, which is labeled as NV, is used to search for the monopole-dipole interaction with nucleons.
    The nucleons are provided by a fused silica half-ball lens, which is labeled as SiO$_2$ .
    The half-ball lens is placed on a tuning fork actuator of an atomic force microscope. A static magnetic field $B_0$ is applied along the symmetry axis of the nitrogen-vacancy center.
    (\textbf{b}) Schematic experimental parameters. The electron spin and the half-ball lens are denoted as $S$ and $M$, respectively.
    The radius of $M$ is $R$.
    $M$ is located right above $S$ and driven to vibrate with amplitude $A$. The distance between $S$ and the bottom of $M$ is $d_0$ when $M$ vibrates to the position nearest $S$.
    (\textbf{c}) Atomic structure and energy levels of the nitrogen-vacancy center in diamond.
    The NV center consists of a substitutional nitrogen atom with an adjacent vacancy cite in the diamond crystal lattice.
    The ground and excited states are denoted as $^3A_2$ and $^3E$. The nitrogen-vacancy center can be excited from $^3A_2$ to $^3E$ by a laser pulse, and decays back to $^3A_2$ emitting photoluminescence.
    The optical transitions are used to initialize and read out the spin state of the nitrogen-vacancy center.
    The spin states $|m_S=0\rangle$ and $|m_S=-1\rangle$ of $^3A_2$ are encoded as a quantum sensor.
    The state of $S$ can be manipulated by microwave pulses.
    }
    \label{Fig1}
\end{figure*}

An NV-based optically detected magnetic resonance setup combined with an atomic force microscope (AFM) \cite{SM} (shown in Fig \ref{Fig1}) is constructed to search for this electron-nucleon interaction
A near-surface electron spin, which is a defect in diamond composed of a substitutional nitrogen atom and a neighboring vacancy cite \cite{PhysRep_NV},  is utilized as a quantum sensor to detect its electron-nucleon interaction with nucleons in a fused silica half-ball lens.
The NV center is less than $10~$ nanometers close to the surface of the diamond, so that it allows close proximity between the electron and the nucleon.
Hereafter, the electron spin of the NV center and the half-ball lens are denoted as $S$ and $M$ for convenience, respectively.
$M$ is placed on a tuning fork actuator of the AFM, which enables us to position $M$ near and away from $S$ as well as to drive $M$ to vibrate with a frequency.
Figure \ref{Fig1}b shows the geometric parameters in the experiment.
The radius of $M$ is $R=250(2.5)~\mu$m.
The vibration amplitude of $M$ is denoted as $A$.
The time-dependent distance between the bottom of $M$ and $S$ can be described as $d=d_0+A[1+\cos(\omega_{\textrm{m}}t)]$, where $d_0$ is the minimal distance between $M$ and $S$, and $\omega_{\textrm{m}}$ is the vibration angular frequency of $M$ driven by the tuning fork.
The effective magnetic field felt by $S$ arising from the hypothetic electron-nucleon interaction can be derived by integrating Eq. \ref{Bsp} over all the nucleons in $M$ as $\vec{B}_{\textrm{eff}}=\hat{r}_{\textrm{c}}B_{\textrm{eff}}$, where $\hat{r}_{\textrm{c}}$ is the unit distance vector along the symmetry axis of $M$ and
\begin{equation}
\label{Beff}
B_{\textrm{eff}}=\frac{\hbar g_s^Ng_p^e\rho}{2m\gamma}f(\lambda, R, d),
\end{equation}
with $\rho=1.33\times10^{30}~$m$^{-3}$ being the number density of nucleons in $M$ and $f(\lambda, R, d) = \lambda [\frac{R}{d+R}e^{-\frac{d}{\lambda}} - e^{-\frac{d+R}{\lambda}} + e^{-\frac{\sqrt{R^2+(d+R)^2}}{\lambda}} + \frac{\lambda\sqrt{R^2+(d+R)^2}}{(d+R)^2}e^{-\frac{\sqrt{R^2+(d+R)^2}}{\lambda}} - \frac{\lambda d}{(d+R)^2}e^{-\frac{d}{\lambda}} + \frac{\lambda^2}{(d+R)^2}e^{-\frac{\sqrt{R^2+(d+R)^2}}{\lambda}} - \frac{\lambda^2}{(d+R)^2}e^{-\frac{d}{\lambda}}]$ (see \cite{SM} for details).
If $M$ is moved far away from $S$ with distance much larger than the force range $\lambda$, the monopole-dipole interaction is negligible.
By comparing the magnetic field detected by $S$ with and without $M$, the electron-nucleon interaction between $S$ and the nucleons in $M$ can be measured.

Figure \ref{Fig1}c shows the atomic structure and energy levels of the NV center.
The ground state of the NV center is an electron spin triplet state $^3A_2$ with three substates $|m_S=0\rangle$ and $|m_S=\pm1\rangle$.
A static magnetic field $B_0$ of about 300~G is applied along the NV symmetry axis to remove the degeneracy of the $|m_S=\pm1\rangle$ spin states.
The spin states $|m_S=0\rangle$ and $|m_S=-1\rangle$ are encoded as a quantum sensor\cite{Arxiv_QuantSensing}.
Microwave pulses with frequency matching the transition between $|m_S = 0\rangle$ and $|m_S = -1\rangle$ are delivered by a copper microwave wire to manipulate the state of the quantum sensor.
The $|m_S=1\rangle$ state remains idle due to the large detuning.
A laser pulse can be applied to pump the NV center from $^3A_2$ to the excited state $^3E$.
When the NV center decays back to $^3A_2$,  photoluminescence can be detected.
The optical process can be utilized to realize state initialization and readout of this quantum sensor.
Because of the convenient state initialization and readout procedures, precise control \cite{NatureCommun_FTGate}, long coherence time \cite{NatureCommun_NVCoherenceTime}, and its atomic size, the NV center serves as a superb magnetic sensor at nanometer scale, which is now extended to search for the axion-mediated interactions beyond the standard model.

If mass $M$ is placed near the electron-spin $S$, a static effective DC magnetic field $B_{eff}$ caused by monopole-dipole interaction will affect $S$.
A straightforward approach to detect such DC magnetic field is to perform a Ramsey sequence \cite{Arxiv_QuantSensing}.
The Ramsey sequence can be written as $\pi /2 - \tau -\pi/2$, where $\pi/2$ stands for the microwave pulse with rotating angle $\pi/2$ and $\tau$ stands for a waiting time. The first $\pi/2$ microwave pulse prepares $S$ to a superposition state $(|0\rangle-i|1\rangle)/\sqrt{2}$.
During the waiting time $\tau$, the electron spin precesses about the z-axis and accumulate a phase proportional to the strength of the magnetic field $B_{eff}$. After the second $\pi/2$ pulse, the phase information will be encoded in the population of the states $|m_S = 0\rangle$, which can be detected with a lase pulse.
However, during the waiting time, noises such as the fluctuation of the Overhouser field and the slow drift of the external static magnetic field, will cause the dephasing effect.
Thus the sensitivity of such method is limited by the dephasing time of the electron spin, which is about $T_2^* = 1~\mu$s measured in our experiment.

\begin{figure}[http]
\centering
\includegraphics[width=1\columnwidth]{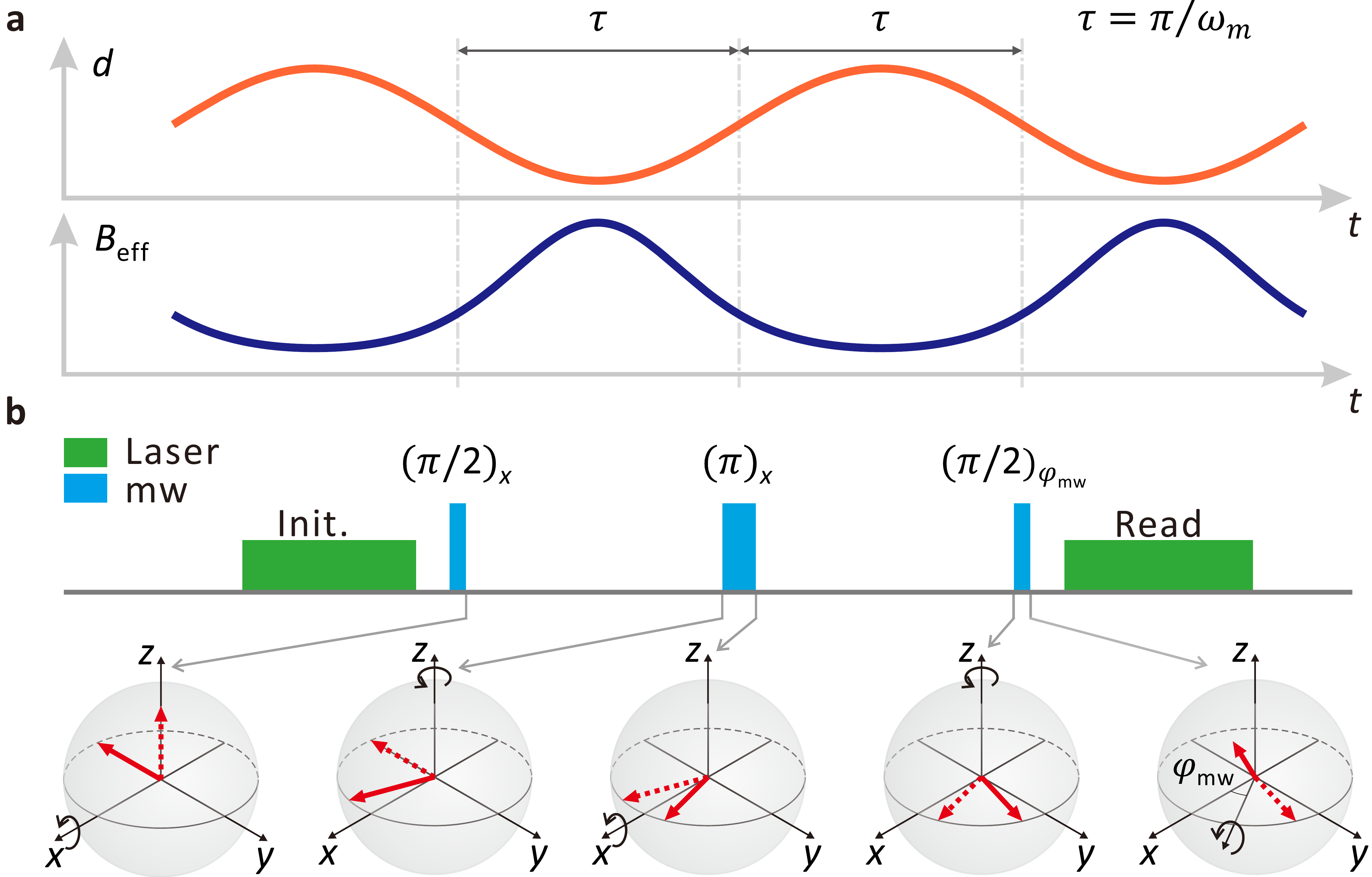}
    \caption{Electron-nucleon interaction detection scheme. (\textbf{a}) Time variation of the distance $d$ (upper) and the effective magnetic field $B_{\textrm{eff}}$ (lower).
    The distance $d$ is between $S$ and the bottom of $M$.
    The waiting time, $\tau=\pi/\omega_\textrm{m}$, is half period of the vibration of $M$, and $B_\textrm{eff}$ is the effective magnetic field on $S$ generated by the nucleons in $M$. (\textbf{b}) Experimental pulse sequence (upper) and state evolution of $S$ (lower). The pulse sequence applied on $S$ is synchronized with the vibration of $M$. Green laser pulses were used to initialize and read the state of S. The microwave $\pi/2$ and $\pi$ pulses were applied only when $M$ passed through the equilibrium point of the vibration.
}
    \label{Fig2}
\end{figure}

To suppress the dephasing effect and to enhance the sensitivity of detecting $B_{eff}$, a spin echo sequence \cite{PR_Echo} can be applied instead of the Ramsey sequence.
The spin echo sequence can be written as $\pi/2 - \tau - \pi -\tau -  \pi/2$.
With this spin echo sequence, the coherence time of the electron spin is enhanced to about $T_2 = 10~\mu$s in our experiment, which is of an order longer than $T_2^*$.
Since the positive phase accumulated during the first waiting time $\tau$ is exactly canceled by the negative phase accumulated during the second $\tau$, the total phase due to static $B_{eff}$ is zero.
To solve this problem, we drive $M$ to vibrate periodically to make $B_{eff}$ an oscillating signal (shown in Fig. \ref{Fig2}a).
If $B_{eff}$ is modulated in-phase with the spin echo sequence, a none-zero accumulated phase due to $B_{eff}$ can be obtained, while the unwanted noise can be canceled.

Figure \ref{Fig2}a shows schematically the distance $d$ and corresponding time-varying effective magnetic field $B_{\textrm{eff}}$ arising from the hypothetical electron-nucleon interaction.
The mass is driven to vibrate with an angular frequency $\omega_{\textrm{m}}=1.18\times10^6~$rad$\cdot$s$^{-1}$.
The vibration amplitude $A$ and shortest distance $d_0$ are $A=41.1(1)~$nm and $d_0=0.5(1)~\mu$m, respectively.
When $M$ vibrates to the position nearest to $S$, the distance $d$ reaches the minimum value $d_0$ and the corresponding effective magnetic field $B_{\textrm{eff}}$ achieves a maximum value.
When $M$ vibrates to the position furthest from $S$, $d$ reaches the maximum value $d_0+2A$ and $B_{\textrm{eff}}$ achieves a minimum value.

Figure \ref{Fig2}b shows the pulse sequence applied on $S$ and the corresponding state evolution of $S$ on the Bloch sphere.
To optimize the phase accumulation, the microwave $\pi/2$ and $\pi$ pulses in the spin echo sequence are applied only when $M$ vibrates passing through the equilibrium point of the vibration.
The electron spin $S$ is initialized into $\ket{m_S = 0}$ by a laser pulse, corresponding to the unit vector along $\hat{z}$ axis in the Bloch sphere.
The first microwave $\pi/2$ pulse transforms the state into $(|0\rangle-i|1\rangle)/\sqrt{2}$.
This $\pi/2$ pulse corresponds to a rotation of the Bloch vector from $\hat{z}$ axis to $-\hat{y}$ axis.
Then $S$ evolves under the effective magnetic field $\vec{B}_{\textrm{eff}}$ for half of the vibration period $\tau$, corresponding to the spin processing around the $\hat{z}$ axis.
As a result, the state is evolved into $(|0\rangle-ie^{i\varphi_0}|1\rangle)/\sqrt{2}$ at the end of the free evolution with corresponding Bloch vector pointing along the axis $\hat{x}\sin\varphi_0-\hat{y}\cos\varphi_0$, where $\varphi_0=\int_{\tau/2}^{3\tau/2}\gamma B_{\textrm{eff}}(t)\cos\theta dt$ is the accumulated phase,
and $\theta=\arccos(1/\sqrt{3})$ is the angle between $\vec{B}_{\textrm{eff}}$ and the NV axis.
The following microwave $\pi$ pulse rotates the Bloch vector by an angle of $\pi$ around $\hat{x}$ axis, transforming the state into $(|0\rangle+ie^{-i\varphi_0}|1\rangle)/\sqrt{2}$.
After the $\pi$ pulse, the electron spin experiences another free evolution for half of the vibration period under $\vec{B}_{\textrm{eff}}$.
At the end of this evolution, the state is evolved into $(|0\rangle+ie^{-i\varphi}|1\rangle)/\sqrt{2}$ with corresponding Bloch vector pointing along the axis $\hat{x}\sin\varphi+\hat{y}\cos\varphi$ in the Bloch sphere picture, with $\varphi=\varphi_0-\int_{3\tau/2}^{5\tau/2}\gamma B_{\textrm{eff}}(t)\cos\theta dt$.
A final microwave $\pi/2$ pulse with phase $\varphi_{\textrm{mw}}$ then rotates the Bloch vector by an angle of $\pi/2$ around the axis $\hat{x}\cos\varphi_{\textrm{mw}}+\hat{y}\sin\varphi_{\textrm{mw}}$, transforming the state into $\cos[(\varphi_{\textrm{mw}}+\varphi)/2]|0\rangle+e^{i\varphi_{\textrm{mw}}}\sin[(\varphi_{\textrm{mw}}+\varphi)/2]|1\rangle$.
After this spin echo sequence, a laser pulse is applied and the photoluminescence intensity $I_{\textrm{PL}}$ is detected.
The measured $I_{\textrm{PL}}$ reflects the population $P_{|0\rangle}$ of state $|m_S = 0\rangle$ for the final state, with $P_{|0\rangle}=1/2+1/2\cos(\varphi_{\textrm{mw}}+\varphi)$. Therefore, $I_{\textrm{PL}}$ can be expressed as a cosine function of $\varphi_{\textrm{mw}}$
\begin{equation}
\label{IPL}
I_{\textrm{PL}}=I_{\textrm{PL},0}+A_{\textrm{PL}}\cos(\varphi_{\textrm{mw}}+\varphi).
\end{equation}
By measuring the photoluminescence intensity $I_{\textrm{PL}}$ with a set of different phases $\varphi_{\textrm{mw}}$ of the final microwave $\pi/2$ pulse, we can extract $\varphi$ which contains the information of $B_{\textrm{eff}}$ arising from the mass-spin interaction.
The coupling $g_s^Ng_p^e$  can be derived to be
\begin{equation}
\label{gsgp}
g_s^Ng_p^e=\frac{1}{\cos\theta}\frac{2m}{\hbar\rho}\frac{\varphi}{\int_{\tau/2}^{3\tau/2}f(\lambda, R, d(t))dt-\int_{3\tau/2}^{5\tau/2}f(\lambda, R, d(t))dt}.
\end{equation}

\begin{figure}[http]
\centering
\includegraphics[width=1\columnwidth]{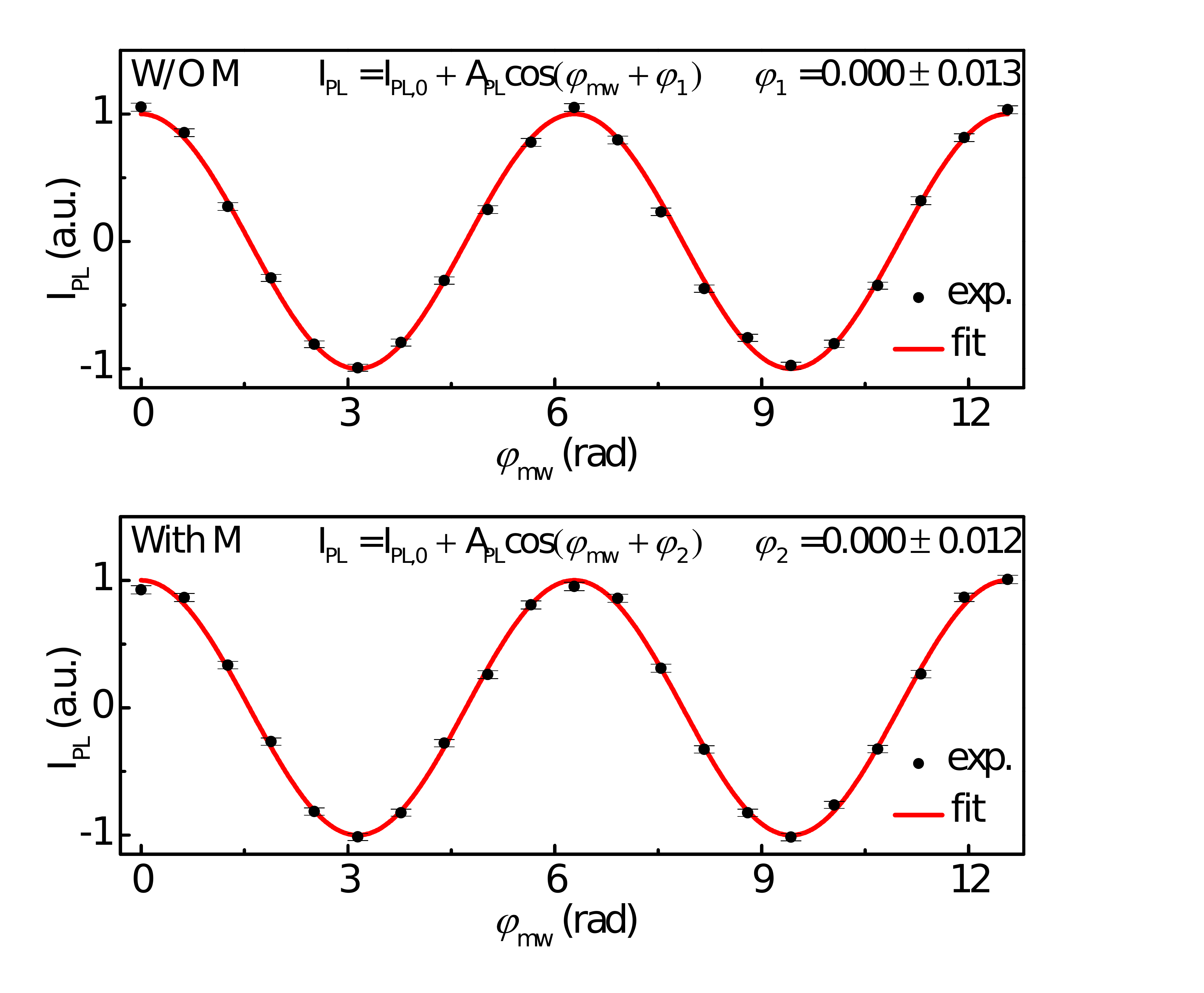}
    \caption{Experimental results for detecting the electron-nucleon interaction. The upper panel shows the measured photoluminescence intensity $I_{\textrm{PL}}$ without $M$, and the lower panel shows that with $M$.
    In both panels, the experimental data are represented by black circles with error bars, and the red solid lines represent the fitting of the experimental data.
    The error bars of the experimental data are the standard deviations.
    The obtained phases $\varphi_1$ and $\varphi_2$ are the accumulated phases of the states of $S$ without and with $M$.
    The phase shift due to the electron-nucleon interaction between $S$ and $M$ is obtained by $\varphi=\varphi_2-\varphi_1$ to be $\varphi=0.000\pm0.018~$rad.
}
    \label{Fig3}
\end{figure}

Figure \ref{Fig3} shows the experimental results.
To exclude the influence of any possible oscillating magnetic field from other sources, we firstly implement the pulse sequence without $M$ as a benchmark experiment.
The experimental data without $M$ is shown in the upper panel of Fig. \ref{Fig3}.
By fitting the data with Eq. \ref{IPL}, we obtain $\varphi_{1}=0.000\pm0.013~$rad as a benchmark.
Then the spin echo sequence is implemented with vibrating $M$ and the result has been shown in the lower panel of Fig. \ref{Fig3}.
The experimental data with $M$ is fitted with Eq. \ref{IPL} to extracted $\varphi_{\textrm{2}}$ with $\varphi_{2}=0.000\pm0.012~$rad.
The accumulated phase $\varphi$ of the electron spin's state owing to $B_{\textrm{eff}}$ generated by $M$, which is obtained by $\varphi=\varphi_{2}-\varphi_{1}$, is determined to be $\varphi=0.000\pm0.018~$rad.
The electron-nucleon interaction has not been observed at the current experimental condition, but an upper limit can be set to constrain the interaction.

\begin{figure}[http]
\centering
\includegraphics[width=1\columnwidth]{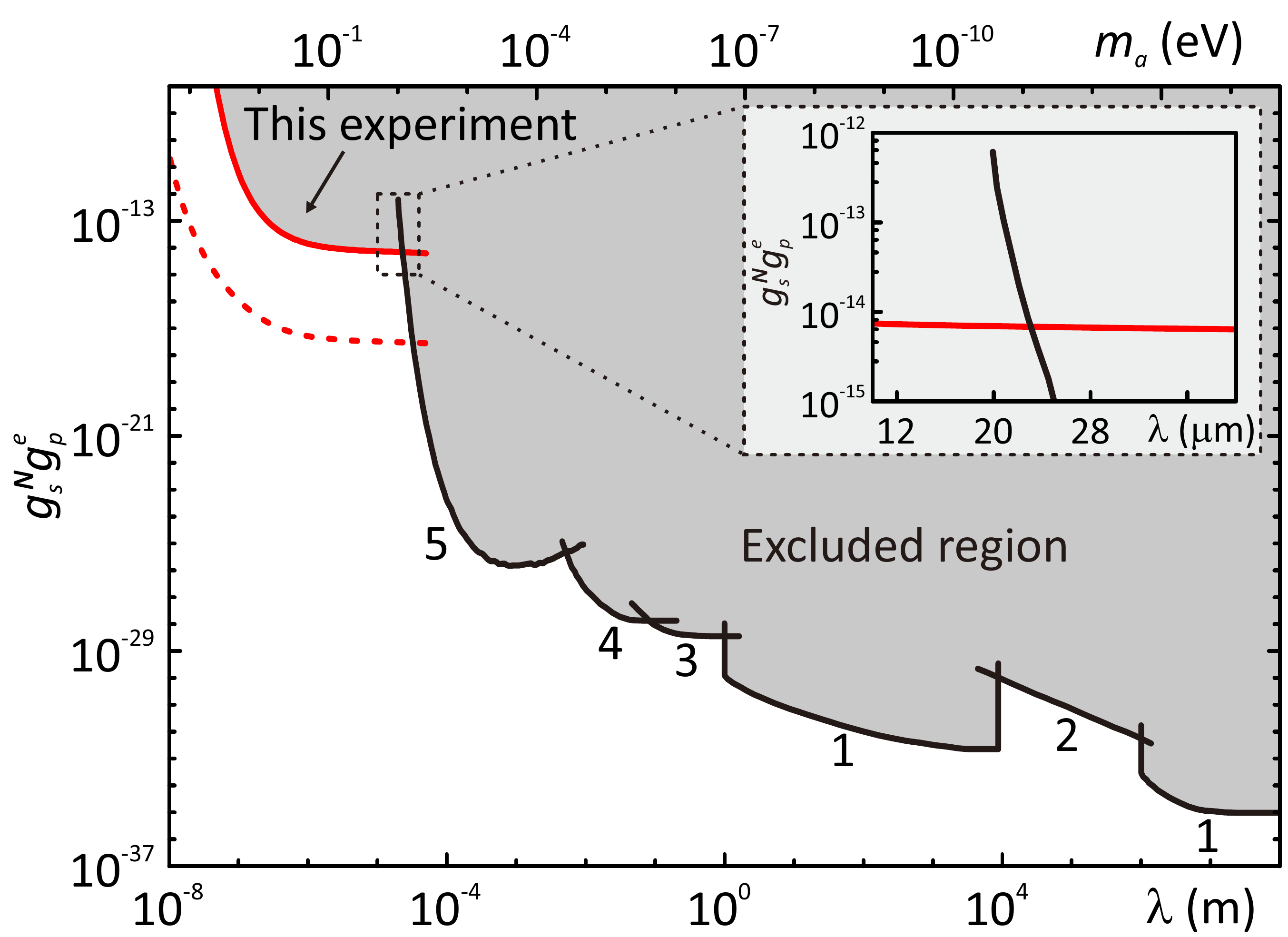}
    \caption{Upper limits on $g_s^Ng_p^e$  as a function of the force range $\lambda$ and mass of ALP $m_a$. Our result is represented as the red solid line. The black solid lines labeled by 1-5 represent the results from references  \cite{PRD78_092006, PRL67_1735, PRL77_2170, PRL82_2439, PRL106_041801}.
    The red dashed line shows the available improvement of the constraint on $g_s^Ng_p^e$ in future \cite{SM}.
    The inset shows a comparison of our result and that from reference \cite{PRL106_041801} with the force range nearby 20~$\mu$m, which illustrates an improvement of two orders more stringent for our result at 20~$\mu$m compared with that from reference \cite{PRL106_041801}.
}
    \label{Fig4}
\end{figure}

Fig.4 shows the new constraint set by this work together with recent constraints from experimental searches for monopole-dipole interations \cite{PRD_LimitCoupling}.
The lines labeled by 1 are from the experiment by Heckel \textit{et al.} \cite{PRD78_092006}, which provides the upper limits in the meter range and above, except a gap from 10~km to 1000~km.
The upper limit in this gap is obtained by the experiment by Wineland \textit{et al.} \cite{PRL67_1735} (solid line labeled by 2).
The experiment by Youdin \textit{et al.} \cite{PRL77_2170} sets the upper limit in the range from 0.08~m to 1~m, which is represented as the solid line labeled by 3.
The solid line labeled by 4 is the upper limit from the experiment by Ni \textit{et al.} \cite{PRL82_2439}, which is in the range from 5~mm to 8~cm.
In the range from 20~$\mu$m to 5~mm, the experiment by Hoedl \textit{et al.} \cite{PRL106_041801} provides the upper limit shown as the solid line labeled by 5.
Our result is represented as the solid red line.
It is derived according to Eq. \ref{gsgp} with $2\delta_{\varphi}$ as an upper bound of $\varphi$, where $\delta_{\varphi}=0.018$ is the standard deviation of the phase accumulation $\varphi$.
Besides $\delta_{\varphi}$, the uncertainties of other experimental parameters such as $d_0$ and $A$ are also taken into account to derive the upper limit \cite{SM}.
From the force range $0.1~\mu m < \lambda < 23 ~\mu m$, our result provided the upper bound for $g_s^Ng_p^e$.
As is shown in the inset of Fig. \ref{Fig4}, the obtained upper bound of the interaction at 20~$\mu$m, $g_s^Ng_p^e < 6.24\times10^{-15}$, is two orders of magnitude more stringent than the bound set by previous experiment \cite{PRL106_041801}.
The possible value of mass of the ALPs,  from $10^{-5}~$eV to $1~$eV (corresponding to a force range $ 0.2~\mu$m $<\lambda<2~$cm), is still allowed by otherwise stringent constraints\cite{PRD_RevPartPhys}.
The unexplored force range left by the previous experiments has now been fully searched in our experiment.

The constraint can be further improved by several strategies in future.
We search for spin-mass interaction by detecting the accumulated phase  of  a single electron spin's state owing to $B_{eff}$.
One effective method is to enhance the coherence time of the electron spin, by synthesizing $^{12}C$ enriched diamond \cite{NatureMater_NVIsotopCoheren} or by applying multi-pulse dynamical decoupling sequences \cite{Nature_DDDu,Science_DDHanson}.
Once the coherence time is prolonged, the ability of detecting the accumulated phase can be enhanced.
On the other hand, the phase accumulation is proportional to the number density of nucleons in the source.
To use materials with high number density of nucleons as the source, such as Bi4Ge3O12 (BGO), can also improve the constraint.
To decrease the measurement uncertainty of the phase accumulation, one can improve the detection efficiency of the photoluminescence and increase the number of experiment scans.
Based on above extensions of techniques, the available constraint, which is shown as the red dashed line in Fig. \ref{Fig4}, could be about 3 orders of magnitude improved from the current result; detailed discussion included in \cite{SM}.

Our platform uses a near-surface NV center together with AFM setup, thus the force range can be focused within micrometers.
The micrometer and submicrometer range, which is not easily accessed in previous experiments, provides a new window for investigating new physics beyond standard model.
The electron-nucleon interaction investigated in our work is one of interactions from new particle exchange \cite{JHighEnergyPhys_MacroForce}.
In future, several related interactions can also be investigated with extension of our method.
For example, spin-spin interaction mediated by APLs, which is recently set a constraint at micrometer scale \cite{PRL_ConstraintDipDip}, can be further explored with submicrometer scale by two coupled NV centers with technologies developed by Grinolds $\textit{et al.}$ \cite{NaturePhys_NanoMI}.
Another case is to explore the interaction mediated by a vector boson, which has been investigated at micrometer force range \cite{PRL_VectorAxial_2013,PRL_VectorAxial_2015}.
Therefore NV centers will not only be a promising quantum sensor for physics within standard model \cite{Nature455_644,Nature455_648,Science347_1129,Science347_1135,Science355_503}, but also be an important platform for searching for new particles predicted by theories beyond the standard model.

We are grateful to H.Y.Yan for his systematic introduction about the spin-dependent forces and fruitful discussion about the experiment.
We thank C.K.Duan and D.J. Kimball for helpful discussion.
We thank L.P.Guo for his help on nitrogen ion implantation.
The fabrication of diamond nanopillars for improving the detection efficiency of the photoluminescence was performed at the USTC Center for Micro and Nanoscale Research and Fabrication.
This work was supported by the National Key Basic Research Program of China (Grants No. 2013CB921800, 2016YFA0502400 and No. 2016YFB0501603), the National Natural Science Foundation of China (Grant No.\ 11227901, No.\ 91636217 and No.\ 31470835) and the Strategic Priority Research Program (B) of the CAS (Grant No. XDB01030400).
J. Du and X. Rong thank financial support by Key Research Program of Frontier Sciences, CAS (Grants No. QYZDY-SSW-SLH004 and QYZDB-SSW-SLH005).
F. Shi and X. Rong thank the Youth Innovation Promotion Association of Chinese Academy of Sciences for the support.
Y.F. Cai is supported in part by the Chinese National Youth Thousand Talents Program, by the CAST Young Elite Scientists Sponsorship Program (2016QNRC001), by the National Natural Science Foundation of China (Nos. 11421303, 11653002), and by the Fundamental Research Funds for the Central Universities.
X. Qin thank support by Fundamental Research Funds for the Central Universities (Grant No. WK2030040081).





\onecolumngrid
\vspace{1.5cm}
\begin{center}
\textbf{\large Supplmentary Material}
\end{center}

\setcounter{figure}{0}
\setcounter{equation}{0}
\makeatletter
\renewcommand{\thefigure}{S\@arabic\c@figure}
\renewcommand{\theequation}{EqS\@arabic\c@equation}
\renewcommand{\bibnumfmt}[1]{[RefS#1]}
\renewcommand{\citenumfont}[1]{RefS#1}

\def\U{\mathrm{U}}
\def\A{\mathcal{A}}
\def\B{\mathcal{B}}
\def\C{\mathcal{C}}
\def\D{\mathcal{D}}
\def\E{\mathcal{E}}
\def\F{\mathcal{F}}
\def\G{\mathcal{G}}
\def\H{\mathcal{H}}
\def\I{\mathcal{I}}
\def\L{\mathcal{L}}
\def\M{\mathcal{M}}
\def\N{\mathcal{N}}
\def\O{\mathcal{O}}
\def\P{\mathcal{P}}
\def\Q{\mathcal{Q}}
\def\T{\mathcal{T}}
\def\R{\mathbb{R}}
\def\J{\mathcal{J}}
\def\S{\mathcal{S}}
\def\K{\mathcal{K}}
\def\Tr{\operatorname{Tr}}

\section {\uppercase\expandafter{\romannumeral1}. Experimental setup}
\begin{figure}[http]
\centering
\includegraphics[width=0.75\columnwidth]{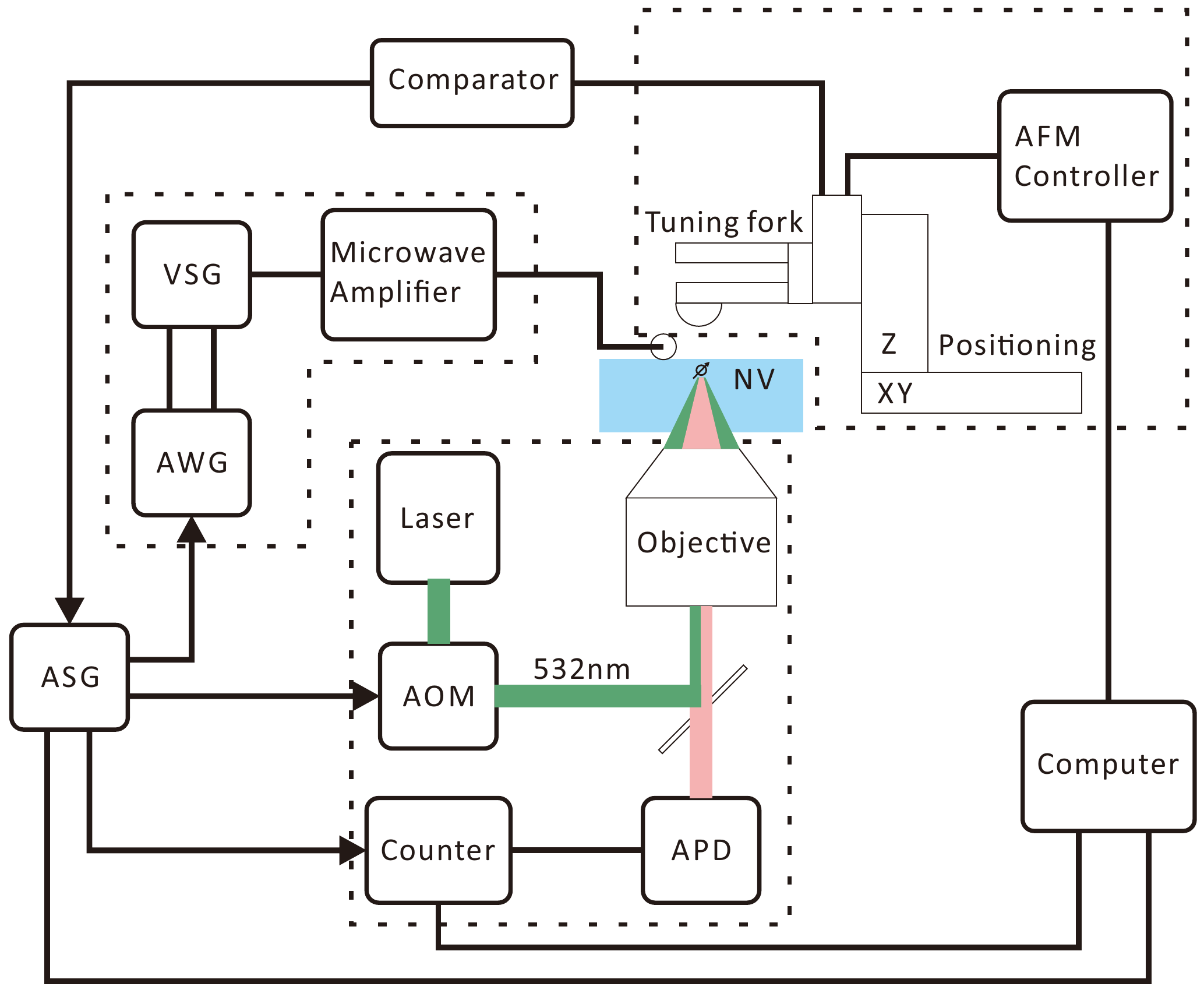}
    \caption{\textbf{Schematic of the experimental setup.} An ODMR setup, which consists of an optical system to initialize and read out the electron spin state of the NV center and a microwave system to manipulate the electron spin state of NV center, and a tuning fork based atomic force microscope to position and drive the half-ball lens, are synchronized by an arbitrary sequence generator (ASG).
}
    \label{FigS1}
\end{figure}

An optically detected magnetic resonance (ODMR) setup combined with an atomic force microscope (AFM) is constructed to investigate the ALP-mediated interaction between an electron spin of a nitrogen-vacancy (NV) center in diamond and a fused silica half-ball lens.
The schematic of the experimental setup is shown in Fig. \ref{FigS1}.

The NV center was created by implantation of 10~keV $\textrm{N}_2^+$ ions into [100] bulk diamond and annealing for 2 hours at $800~^{\circ}$C in vacuum.
The implantation depth is estimated to be 5-11 nm below the diamond surface\cite{RF1}.
Then the diamond was oxidatively etched for 4 hours at $580~^{\circ}$C.
After oxidative etching, the depth of the NV center is estimated to be $1$--$7$ nm\cite{RF2}.
The diamond was installed in an ODMR setup.
The ODMR setup consists of optical and microwave systems.
The optical system enables state initialization and readout of the electron spin via a 532 nm laser pulse.
The laser pulse passed through an acousto-optic modulator (AOM, ISOMET 1250C) and an objective (Olympus, LUCPLFLN 60X, NA 0.7) before being focused on the NV center.
The AOM was passed through twice by the laser pulse to preserve the longitudinal relaxation time of the NV center from laser leakage effects.
When a 532 nm laser pulse is applied, phonon sideband fluorescence with wavelength of $650$--$800$ nm can be detected to determine the state of the electron spin.
The fluorescence went through the same objective and was collected by an avalanche photodiode (Perkin Elmer SPCM-AQRH-14) with a counter card.
To increase the fluorescence collection efficiency, nanopillars has been fabricated on the diamond by electron beam lithography (EBL) and reactive ion etching (RIE).
The microwave system enables state manipulation of the electron spin.
Microwave pulses were generated by IQ modulation.
The IQ baseband was provided by a 4.2 GSa/s arbitrary waveform generator (AWG, Keysight 81180A).
A vector signal generator (VSG, Keysight E8267D) was used as the IQ modulator and also provided the carrier frequency.
The phase and amplitude balance of the modulator were carefully calibrated with a signal analyzer (Keysight N9020A).
The generated microwave pulses were amplified by a power amplifier (Mini-Circuits ZHL-16W-43-S+) and delivered by a copper microwave wire with a diameter of $20~\mu$m to manipulate the electron spin state.

The fused silica half-ball lens, which is denoted as $M$ in the main text and hereafter, was supplied by Edmund Optics Inc.
According to the datasheet, the radius of $M$ is $250(2.5)~\mu$m.
The surface roughness of $M$ is estimated to be $0.01~\mu$m by measurements with AFM.
$M$ can be positioned near and away from the NV center by an AFM system.
The AFM system is composed of a positioning system and a tuning fork.
The positioning system is used to determine and change the location of $M$.
The tuning fork is utilized to drive $M$ to vibrate.
The vibration angular frequency and amplitude can be controlled from the software of the AFM controller (Asylum Research MFP3D).
The angular frequency is set to be $\omega_\textrm{m}=1.18\times10^6~$rad$\cdot$s$^{-1}$ to match a natural frequency of the tuning fork.
The vibration amplitude depends on the driving voltage set from the software.
By analyzing the noise spectrum of the tuning fork, a coefficient of $6.85~$nm/V is obtained.
The driving voltage is set to $6.00~$V.
Therefore, the vibration amplitude of $M$ is estimated to be $A=41.1(1)~$nm, where the uncertainty is estimated from the precision of the coefficient and driving voltage.

The state initialization, manipulation, readout of the electron spin, and the vibration of $M$, are synchronized by an arbitrary sequence generator (ASG, Quantum Precision Device Co. ASG-GT50-C).
An oscillating electronic signal in phase with the vibration is generated due to the piezoelectric effect of quartz tuning fork.
The oscillating signal is transformed into a periodic rectangular pulse train by a comparator.
The ASG can be triggered by the rectangular pulse train to generate rectangular pulses, which are used to trigger or control the AWG, counter card, and the AOM.

\section {\uppercase\expandafter{\romannumeral2}. Effective magnetic field from the half-ball lens}
The effective magnetic field arising from the hypothetical monopole-dipole interaction between the electron spin and a nucleon is
\begin{equation}
\label{SM_Bsp}
\vec{B}_{\textrm{sp}}(\vec{r})=\frac{\hbar g_s^Ng_p^e}{4\pi m\gamma}(\frac{1}{\lambda r}+\frac{1}{r^2})e^{-\frac{r}{\lambda}}\hat{r}.
\end{equation}
The effective magnetic field from the half-ball lens $M$ can be derived by integrating $\vec{B}_{\textrm{sp}}(\vec{r})$ over all the nucleons in $M$,
\begin{equation}
\label{SM_vecBeff_Int}
\vec{B}_{\textrm{eff}}=\int_V\vec{B}_{\textrm{sp}}(\vec{r})\rho dV.
\end{equation}
The component of $\vec{B}_{\textrm{eff}}$ perpendicular to the symmetry axis of $M$ is zero by symmetry.
Therefore, $\vec{B}_{\textrm{eff}}$ can be written as
\begin{equation}
\label{SM_vecBeff}
\vec{B}_{\textrm{eff}}=\hat{r}_{\textrm{c}}B_{\textrm{eff}},
\end{equation}
where $\hat{r}_{\textrm{c}}$ is the unit distance vector along the symmetry axis of $M$.
$B_{\textrm{eff}}$ can be derived by integrating the component of $\vec{B}_{\textrm{sp}}(\vec{r})$ along the symmetry axis of $M$ over all the nucleons in $M$,
\begin{equation}
\label{SM_Beff_Int}
B_{\textrm{eff}}=\int_V\frac{\hbar g_s^Ng_p^e}{4\pi m\gamma}(\frac{1}{\lambda r}+\frac{1}{r^2})e^{-\frac{r}{\lambda}}\frac{z}{r}\rho dV,
\end{equation}
where $z$ is the component of $\vec{r}$ along the symmetry axis.
The integration can be calculated in a cylindrical coordinate system.
The volume element $dV$ is written as
\begin{equation}
\label{SM_dV}
dV=ldld\phi dz,
\end{equation}
where $l$ is the radial distance and $\phi$ is the azimuth.
The distance $r$ can be described with the cylindrical coordinate as
\begin{equation}
\label{SM_r}
r=\sqrt{z^2+l^2},
\end{equation}
By substituting Eq. \ref{SM_dV} and \ref{SM_r} into Eq. \ref{SM_Beff_Int}, we get
\begin{equation}
\label{SM_Beff_Int_cyl}
B_{\textrm{eff}}=\int_d^{d+R}dz\int_0^{\sqrt{R^2-(d+R-z)^2}}dl\int_0^{2\pi}d\phi[\frac{\hbar g_s^Ng_p^e}{4\pi m\gamma}(\frac{1}{\lambda \sqrt{z^2+l^2}}+\frac{1}{z^2+l^2})e^{-\frac{\sqrt{z^2+l^2}}{\lambda}}\frac{z}{\sqrt{z^2+l^2}}\rho l].
\end{equation}
The integration in Eq. \ref{SM_Beff_Int_cyl} is completed to derive $B_{\textrm{eff}}$,
\begin{equation}
\label{SM_Beff}
B_{\textrm{eff}}=\frac{\hbar g_s^Ng_p^e\rho}{2m\gamma}f(\lambda, R, d),
\end{equation}
with
\begin{equation}
\label{SM_f}
\begin{aligned}
f(\lambda, R, d) = \lambda [&\frac{R}{d+R}e^{-\frac{d}{\lambda}} - e^{-\frac{d+R}{\lambda}} + e^{-\frac{\sqrt{R^2+(d+R)^2}}{\lambda}} + \frac{\lambda\sqrt{R^2+(d+R)^2}}{(d+R)^2}e^{-\frac{\sqrt{R^2+(d+R)^2}}{\lambda}} \\
&- \frac{\lambda d}{(d+R)^2}e^{-\frac{d}{\lambda}} + \frac{\lambda^2}{(d+R)^2}e^{-\frac{\sqrt{R^2+(d+R)^2}}{\lambda}} - \frac{\lambda^2}{(d+R)^2}e^{-\frac{d}{\lambda}}].
\end{aligned}
\end{equation}

\section {\uppercase\expandafter{\romannumeral3}. Upper bound of $g_s^Ng_p^e$}
According to the state evolution, the final state of the electron spin after the spin echo sequence is $\cos[(\varphi_{\textrm{mw}}+\varphi)/2]|0\rangle+e^{i\varphi_{\textrm{mw}}}\sin[(\varphi_{\textrm{mw}}+\varphi)/2]|1\rangle$, with
\begin{equation}
\label{SM_varphi}
\varphi=\int_{\tau/2}^{3\tau/2}\gamma B_{\textrm{eff}}\cos\theta dt-\int_{3\tau/2}^{5\tau/2}\gamma B_{\textrm{eff}}\cos\theta dt.
\end{equation}
We search for the hypothetical monopole-dipole interaction, $g_s^Ng_p^e$, by measuring $\varphi$.
According to Eq. \ref{SM_Beff} and \ref{SM_varphi}, the coupling $g_s^Ng_p^e$ can be derived as
\begin{equation}
\label{SM_gsgp_origin}
g_s^Ng_p^e=\frac{1}{\cos\theta}\frac{2m}{\hbar\rho}\frac{\varphi}{\int_{\tau/2}^{3\tau/2}f(\lambda, R, d)dt-\int_{3\tau/2}^{5\tau/2}f(\lambda, R, d)dt},
\end{equation}
where
\begin{equation}
\label{SM_d}
d=d_0+A[1+\cos(\omega_{\textrm{m}}t)]
\end{equation}
describes the vibration of $M$.
The experimental result gives $\varphi=0\pm0.018~$rad.
Therefore, the effect of the hypothetical monopole-dipole interaction is not observed, but an upper bound of $g_s^Ng_p^e$ can be set according to the uncertainties of $\varphi$ and the experimental parameters such as $d_0$ and $A$.

To derive the upper bound of $g_s^Ng_p^e$ with the uncertainties of the experimental parameters taken into consideration, we rewrite Eq. \ref{SM_gsgp_origin} as
\begin{equation}
\label{SM_gsgp}
g_s^Ng_p^e=\frac{\varphi}{h(\lambda; R, d_0, A, \theta)},
\end{equation}
with
\begin{equation}
\label{SM_hfunc}
h(\lambda; R, d_0, A, \theta)=\frac{\hbar\rho\cos\theta}{2m}[\int_{\tau/2}^{3\tau/2}f(\lambda, R, d)dt-\int_{3\tau/2}^{5\tau/2}f(\lambda, R, d)dt].
\end{equation}
The upper bound of $g_s^Ng_p^e$ is calculated as
\begin{equation}
\label{SM_sup_gsgp}
\sup(g_s^Ng_p^e)=\frac{\sup(\varphi)}{\min[h(\lambda; R, d_0, A, \theta)]},
\end{equation}
where $\sup(\varphi)$ represents the upper bound of $\varphi$, and $\min[h(\lambda; R, d_0, A, \theta)]$ is the minimum value of $h(\lambda; R, d_0, A, \theta)$.
We take $\sup(\varphi)=0.036~$rad, two times of the uncertainty of the measured $\varphi$.
The minimum value of $h(\lambda; R, d_0, A, \theta)$ is numerically calculated with the parameters $R$, $d_0$, $A$, and $\theta$ taken within the uncertainty ranges.
The derived upper bound of $g_s^Ng_p^e$ is shown as the red solid line in Fig. 4 in the main text.

The result can be further improved.
Since the sensitivity for detecting the effective magnetic field is limited by the coherence time of the electron spin, a method to improve the bound is to prolong the coherence time.
The coherence time of the near surface NV center in our experiment is prolonged from $T_2^*=1~\mu$s to $T_2=10~\mu$s with a spin echo sequence.
By applying multi-pulse dynamical decoupling sequences such as a CPMG sequence, the coherence time can be further prolonged.
Recently, a coherence time of $T_{2, \textrm{CPMG1024}}=1280~\mu$s is reported for near-surface NV centers with a CPMG-1024 sequence applied \cite{PRL_NVT2CPMG1024}.
The vibration of $M$ can be controlled to match the CPMG sequence.
For example, $M$ can be driven to vibrate with an angular frequency of $\omega_{\textrm{m}}=K\pi/T_{2, \textrm{CPMG1024}}=2.51\times10^6~$rad$\cdot$s$^{-1}$, where $K=1024$ is the number of $\pi$ pulses.
The CPMG-1024 sequence is synchronized with the vibration of $M$ such that the $\pi$ pulses are applied only when $M$ is passing through the equilibrium point of the vibration.
In this case, Eq. \ref{SM_gsgp} still holds as long as Eq. \ref{SM_hfunc} is replaced by
\begin{equation}
\label{SM_hfunc_CPMGK}
h(\lambda; R, d_0, A, \theta)=\frac{\hbar K\rho\cos\theta}{4m}[\int_{\tau/2}^{3\tau/2}f(\lambda, R, d)dt-\int_{3\tau/2}^{5\tau/2}f(\lambda, R, d)dt].
\end{equation}

According to Eq. \ref{SM_gsgp}, strategies to reduce the uncertainty of $\varphi$ and those to maximize $h(\lambda; R, d_0, A, \theta)$ can be taken to improve the result.
A straightforward way to increase $h(\lambda; R, d_0, A, \theta)$ is to use materials with higher $\rho$, such as Bi4Ge3O12 (BGO), as the nucleon source.
The number density of nucleons in BGO is $\rho=4.29\times10^{30}~$m$^{-3}$.
Another method is to optimize the geometry parameters such as $d_0$ and $A$ to increase $h(\lambda; R, d_0, A, \theta)$.
The distance $d_0$ can be set to about $100~$nm, and the amplitude of the vibration can be improved to $A=400~$nm.
To reduce the uncertainty of $\varphi$, we can improve the detection efficiency of the photoluminescence and increase the number of experiment scans.
The detection efficiency can be improved to achieve a photoluminescence rate of $1.7\times10^6~$counts/s \cite{NanoLetter_PLrate}, which is 17 times larger than that in the current experiment.
Considering these improvements, the estimated available upper bound of $g_s^Ng_p^e$ is shown as the red dashed line in Fig. 4 in the main text, which is about 3 orders of magnitude more stringent than than that from the current experiment result.


\begin{thebibliography}{99}
\bibitem{SM} Materials and methods are avaiable as supporting material.
\end{thebibliography}
\end{document}